\UseRawInputEncoding
\documentclass[conference]{IEEEtran}
\IEEEoverridecommandlockouts
\usepackage{listings}
\usepackage{cite}
\usepackage{amsmath,amssymb,amsfonts}
\usepackage[font=footnotesize, labelfont=bf, justification=centering]{caption}
\usepackage{algorithmic}
\usepackage{enumitem}
\usepackage{multirow}
\usepackage{pifont}
\usepackage[utf8]{inputenc}
\usepackage{listingsutf8}
\usepackage{booktabs}
\usepackage{graphicx} 
\usepackage{makecell}   
\usepackage{xcolor}
\begin{document}

\title{AssertCoder: LLM-Based Assertion Generation via Multimodal Specification Extraction\\
}

\author{
\begin{minipage}[t]{0.45\textwidth}
  \centering
  \textbf{Enyuan Tian}\\
  \textit{Institute of Software, Chinese Academy of Sciences}\\
  \textit{University of Chinese Academy of Sciences}\\
  Beijing, China\\
  tianenyuan22@mails.ucas.ac.cn
\end{minipage}
\hfill
\begin{minipage}[t]{0.45\textwidth}
  \centering
  \textbf{Yiwei Ci, Qiusong Yang\textsuperscript{*}}\\
  \textit{Institute of Software, Chinese Academy of Sciences
}\\
  Beijing, China\\
  \{yiwei, qiusong\}@iscas.ac.cn
\end{minipage}
\\[1.5em]
\begin{minipage}[t]{0.45\textwidth}
  \centering
  \textbf{Yufeng Li}\\
  \textit{Institute of Computing Technology, Chinese Academy of Sciences
}\\
  Beijing, China\\
  crazybinary494@gmail.com
\end{minipage}
\hfill
\begin{minipage}[t]{0.45\textwidth}
  \centering
  \textbf{Zhichao Lyu}\\
  \textit{Institute of Software, Chinese Academy of Sciences
}\\
  Beijing, China\\
  lvzhichao18@mails.ucas.ac.cn
\end{minipage}
}
\maketitle

\begingroup
\renewcommand\thefootnote{}\footnotetext{
    \fontsize{8pt}{8pt}\selectfont
    \textsuperscript{*}Corresponding author: Qiusong Yang
}
\endgroup
\renewcommand\thefootnote{\arabic{footnote}}

\begin{abstract}
Assertion-Based Verification (ABV) is critical for ensuring functional correctness in modern hardware systems. However, manually writing high-quality SVAs remains labor-intensive and error-prone. To bridge this gap, we propose AssertCoder, a novel unified framework that automatically generates high-quality SVAs directly from multimodal hardware design specifications. AssertCoder employs a modality-sensitive preprocessing to parse heterogeneous specification formats (text, tables, diagrams, and formulas), followed by a set of dedicated semantic analyzers that extract structured representations aligned with signal-level semantics. These representations are utilized to drive assertion synthesis via multi-step chain-of-thought (CoT) prompting. The framework incorporates a mutation-based evaluation approach to assess assertion quality via model checking and further refine the generated assertions. Experimental evaluation across three real-world Register-Transfer Level (RTL) designs demonstrates AssertCoder's superior performance, achieving an average increase of 8.4\% in functional correctness and 5.8\% in mutation detection compared to existing state-of-the-art approaches.
\end{abstract}

\begin{IEEEkeywords}
Hardware Formal Verification, Assertion Generation, Large Language Models.
\end{IEEEkeywords}

\vspace{-0.5em}
\section{Introduction}
\vspace{-0.5em}
Modern system-on-chip (SoC) designs require rigorous functional verification to ensure correctness under complex control logic, aggressive timing constraints, and highly interconnected subsystems. Assertion-Based Verification (ABV), a cornerstone of formal verification, addresses this need by embedding behavioral specifications as SVAs~\cite{bergeron2007writing}. These assertions act as temporal monitors, enabling early bug detection through both simulation and formal analysis. However, manual SVA development remains a scalability bottleneck due to its labor-intensive nature and reliance on domain expertise~\cite{witharana2022survey}.

To address this challenge, researchers investigated methods based on static analysis\cite{orenes2021autosva,fang2023rmap},
dynamic data mining\cite{danese2017ateam,germiniani2022harm,vasudevan2010goldmine} and natural language processing\cite{aditi2022hybrid,frederiksen2020automated,harris2016glast,keszocze2019chatbot,parthasarathy2021spectosva}, aiming to extract assertions directly
from specification. A key challenge stems from the semantic gap between semi-structured specifications and the formal temporal logic required for assertions\cite{chang2024natural}. While specifications are multimodal, their translation into precise, semantically consistent SVAs lacks systematic methodologies, often leading to semantic mapping errors.

With the recent advances in Large Language Model (LLM), researchers apply LLM to generate assertion from specifications. Some attempts\cite{yan2025assertllm,mali2024chiraag,sun2023towards,kande2024security} primarily explored few-shot and zero-shot LLM prompting to generate SVAs from high-level specifications. Although they are suitable for low-complexity specifications, such methods often struggle to generalize across the wide spectrum of specification formats, which include tabular structures, architectural diagrams, and symbolic timing constraints. Other approaches, such as \cite{bai2025assertionforge,wu2025spec2assertion,pulavarthi2025llms}, adopt a single generation paradigm without incorporating semantic feedback, limiting their ability to handle ambiguity and ensure assertion completeness.  

To address these limitations, we introduce \textbf{AssertCoder}, a unified framework for synthesizing SVAs from multimodal hardware design specifications. AssertCoder initiates with modality-aware preprocessing and dedicated analyzers that extract structured, signal-level information from diverse specifications (text, tables, diagrams, and formulas). These are consolidated into a unified intermediate representation, enabling assertion generation via a multi-step CoT prompting strategy~\cite{wei2022chain}. To ensure the quality of SVAs, AssertCoder incorporates a mutation-guided evaluation and refinement with model checking\cite{clarke1997model}, allowing regeneration of assertions in response to undetected mutation iteratively. This integration of semantic abstraction, structured prompting, and feedback-guided refinement enables high-quality assertion generation across real-world RTL designs.

Our main contributions are as follows:

\begin{table*}[t]
\scriptsize
\centering
\caption{COMPARISONS OF LLM-BASED ASSERTION GENERATION METHODS}
\label{tab:comparison}
\resizebox{\textwidth}{!}{%
\begin{tabular}{|c|c|c|c|c|c|c|c|}
\hline
\thead{Method/Citation} & \thead{Framework\\Input} & \thead{SPEC\\Modality} & \thead{Main\\Prompt Strategy} & \thead{RAG\\Integration} & \thead{Verification\\Target} & \thead{Testbench\\Scale} & \thead{Iterative\\Refinement} \\ \hline
\makecell{AssertLLM\cite{yan2025assertllm}\\(Yan et al., 2024)} & \makecell{SPEC} & \makecell{Multi\\(only Waveform Diagram)} & \makecell{Template-based Structured} & \makecell{\ding{51}\\(SVA Textbooks)} & \makecell{Functional} & \makecell{Mid-scale} & \ding{55} \\ \hline
\makecell{AssertionForge\cite{bai2025assertionforge}\\(Bai et al., 2025)} & \makecell{RTL + SPEC} & Text-only & \makecell{Task Decomposition} &\makecell{\ding{51}\\(Garph RAG)} & \makecell{Functional} & \makecell{Large-scale} & \ding{55} \\ \hline
\makecell{Spec2Assertion\cite{wu2025spec2assertion}\\(Wu et al., 2025)} & SPEC & Text-only & \makecell{Progressive Regularization CoT} & \ding{55} & \makecell{Functional} & \makecell{Mid-scale} & \makecell{\ding{55}} \\ \hline
\makecell{Assertion by LLMs\cite{kande2024security}\\(Kande et al., 2024)} & \makecell{RTL + SPEC} & Text-only & \makecell{Few-shot with Structured Examples} & \ding{55} & \makecell{Security} & \makecell{Mid-scale} & \makecell{\ding{51}\\(Manual)} \\ \hline
\makecell{ChIRAAG\cite{mali2024chiraag}\\(Mali et al., 2024)} & \makecell{RTL + SPEC} & Text-only & \makecell{Structured Prompting via JSON} & \ding{55} & \makecell{Functional} & \makecell{Mid-scale} & \makecell{\ding{51}\\(Partial\textsuperscript{\textdagger})} \\ \hline
\makecell{NL2SVA\cite{sun2023towards}\\(Sun et al., 2024)} & \makecell{RTL + SPEC} & Text-only & \makecell{Circuit-aware CoT} & \ding{55} & \makecell{Functional} & \makecell{Small-scale} & \makecell{\ding{51}\\(Manual)} \\ \hline
\makecell{AssertionLLM\cite{pulavarthi2025llms}\\(Pulavarthi et al., 2025)} & \makecell{RTL + SPEC} & Text-only & \makecell{Instruction-style Prompting} & \ding{55} & \makecell{Functional} & \makecell{Mid-scale} & \ding{55} \\ \hline
\textbf{\makecell{AssertCoder\\(Our Work)}} & \textbf{\makecell{SPEC}} & \textbf{Multi\textsuperscript{*}} & \textbf{\makecell{Alignment-aware Multi-step CoT}} & \textbf{\makecell{\ding{51}\\(Design-aware Retrieval)}} & \textbf{\makecell{Functional}} & \textbf{\makecell{Full-scale}} & \textbf{\makecell{\ding{51}\\(Automatic)}} \\ \hline
\end{tabular}%
}
\vspace{2pt} 
\parbox{\textwidth}{\scriptsize \textsuperscript{*}Multimodal specification includes state diagrams, dataflow diagrams, architecture blocks, etc., all transferred into pseudo-code or natural language to enhance LLM understanding.}
\parbox{\textwidth}{\scriptsize \textsuperscript{\textdagger}ChIRAAG performs partial automated refinement, handling only syntax, basic timing, and missing signal errors; most assertion and design issues still require manual intervention.}
\vspace{-3em}
\end{table*}
	1.	We propose a modality-sensitive classification and analysis mechanism that dispatches heterogeneous specification content (text, tables, diagrams, and formulas) to dedicated analyzers. Each analyzer extracts structured semantic representations, which are converted into a unified information specification to support the SVA generation task.

	2.	We develop a semantics-driven, retrieval-augmented assertion synthesis engine that converts structured specification into formal properties via a four-step process: unified semantic decomposition, pattern selection, temporal binding, and SVA syntax synthesis. The method dynamically selects and instantiates assertion templates using timing and signal semantics.
    
	3.	We adopt a mutation-based approach to evaluate the functionality covered by assertions, aiming to enhance their quality. To comprehensively assess assertion coverage, we employ model checking. This process involves refining assertions to verify functionality with undetected mutations. As a result, potential bugs can be identified through these enhanced assertions. 
    
	4.	We implement and evaluate the framework on 3 real-world RTL modules with multimodal specifications. The results show that the SVA generated by our framework, compared to the baseline, not only generate a greater number of syntax correctness SVAs, but also increases the functional correctness by 8.4\% and the mutation detection by 5.8\%.
\vspace{-0.5em} 
\section{Related Work}
\vspace{-0.5em}
Prior efforts in assertion generation span a spectrum of methodologies, ranging from RTL-guided methods to specification-based approaches. Table~\ref{tab:comparison} provides a comparative overview of representative works.

Recent works, such as AssertionForge\cite{bai2025assertionforge}, NL2SVA\cite{sun2023towards}, and AssertionLLM\cite{pulavarthi2025llms}, rely on both RTL and specification as input. While RTL provides refined temporal semantics and signal-level observability, these methods risk semantic drift if the design under test (DUT) contains  bugs. During the generation process, reliance on these error information inevitably has an impact on SVA, leading to degraded assertion quality.

In contrast, Spec2Assertion\cite{wu2025spec2assertion} and AssertLLM\cite{yan2025assertllm} operate solely on textual or waveform-based specifications. However, these systems do not fully leverage the multimodal abundance of design documents (tables, timing diagrams, and architectural diagrams) and thus often produce assertion with limited functional coverage.

Compared to these methods, our approach focuses on comprehensive modality semantic extraction, structured prompting, and mutation-guided refinement, without RTL information. As summarized in Table~\ref{tab:comparison}, it combines multimodal specification extraction through alignment-aware, multi-step CoT reasoning with automatic assertion refinement.
\section{The AssertCoder Framework}
\subsection{Overview}
\begin{figure*}[t]
    \centering
    \includegraphics[width=0.95\textwidth]{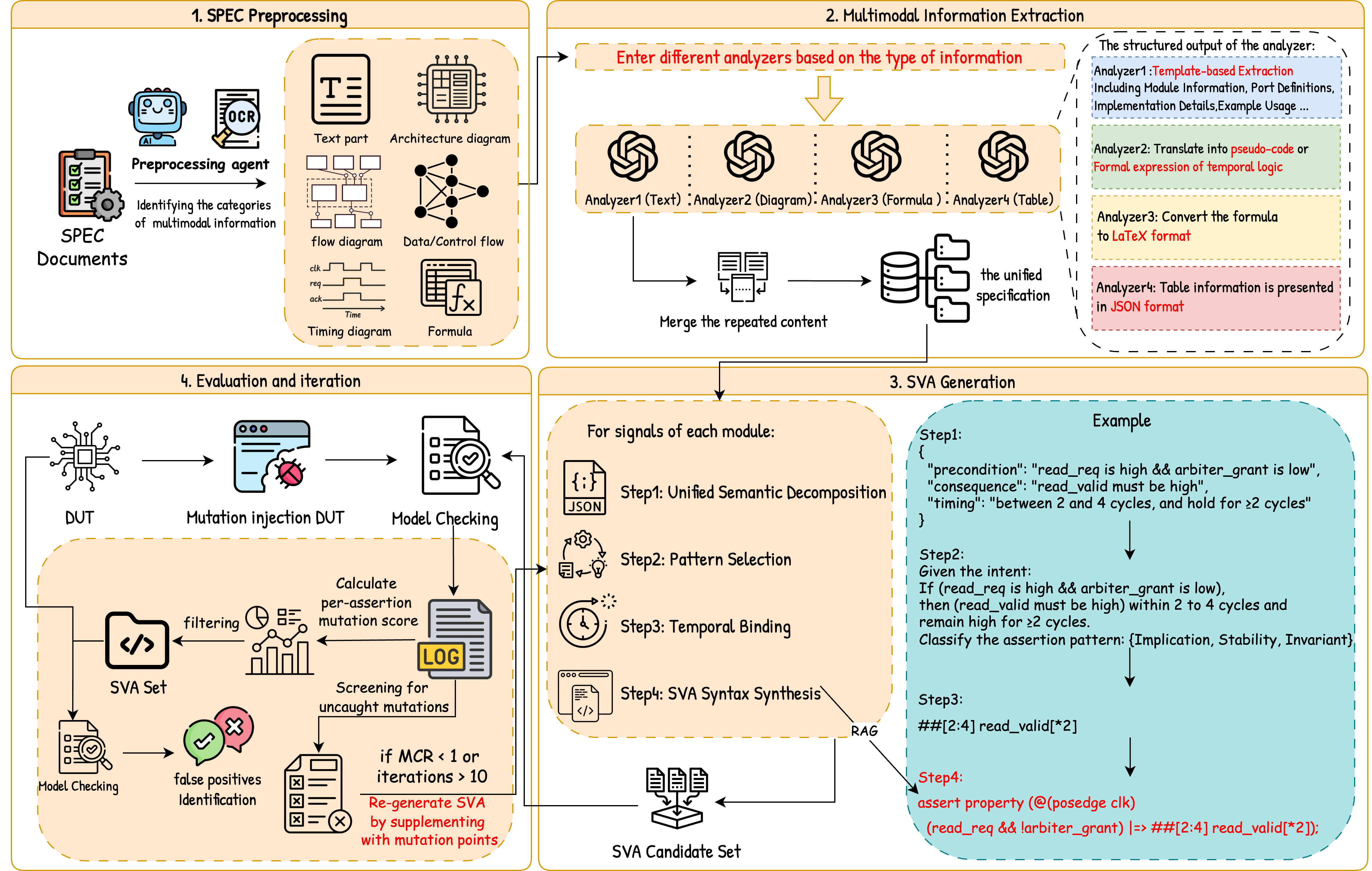}
    \caption{The Overview of AssertCoder Framework}
    \label{fig:Overview}
    \vspace{-2em}
\end{figure*}

The AssertCoder framework is designed to generate SVAs from multimodal specifications. As shown in Fig.~\ref{fig:Overview}, AssertCoder decomposes the overall process into four stages: (1) \textbf{SPEC Preprocessing}, (2) \textbf{Multimodal Information Extraction}, (3) \textbf{SVA Generation}, and (4) \textbf{Evaluation and Iteration}. It integrates a modality-sensitive analysis, multi-step CoT prompting, semantic alignment and retrieval-augmented generation (RAG) to transform specifications into SVAs. To ensure high-quality assertion generation, the framework incorporates a mutation-based mechanism that systematically evaluates the effectiveness of assertion sets through model checking and guides iterative refinement through feedback. 

\subsection{SPEC Preprocessing}
To robustly process specifications, often delivered as PDFs containing text, tables, diagrams, and formulas, AssertCoder initiates with a preprocessing stage that segments the document into structurally meaningful blocks. This step is essential for enabling modality-sensitive analysis in later stages.

We adopt a hybrid approach combining classical document parsing and LLM-based semantic classification. The PDF is first parsed using LayoutParser\cite{shen2021layoutparser}, a deep learning layout analyzer pretrained on PubLayNet\cite{zhong2019publaynet}, to extract visual elements such as paragraphs, tables, and diagrams, along with bounding boxes, content type hints, and position information.

The output of this parser ensemble is a flattened token stream composed of content blocks, each enriched with layout-derived information. These candidate blocks are passed to an LLM via prompting strategy for semantic classification. For each block, the LLM outputs two key attributes: the modality (e.g., TEXT, TABLE, FORMULA, DIAGRAM) and the semantic category (e.g., Architecture, Timing Behavior, Control Logic, Reset Behavior, Module Interface, etc.). The prompt includes structured input-output examples and explicitly instructs the model to reason about signal semantics, control logic, and structural content. An example of the full prompt structure is shown in Listing~\ref{lst:prompt1}.

\begin{lstlisting}[caption={Prompt Template for SPEC Preprocessing Agent}, label={lst:prompt1}]
(*@\textbf{\textcolor{blue}{SYSTEM ROLE}}@*)
You are a hardware specification analyst with expertise in signal semantics, interface conventions, and timing behavior. You are assisting in classifying document blocks extracted from layout-aware parsers.
(*@\textbf{\textcolor{blue}{TASK DESCRIPTION}}@*)
Given a structured content block from a hardware design specification extracted using a layout-aware parser identify:
(*@\textbullet{}@*) The modality of the block: {TEXT,TABLE,FORMULA,DIAGRAM}
(*@\textbullet{}@*) Assign a semantic category from the following(including but not limited to):
   - Architecture      - Module/Interface Declaration
   - Timing Behavior   - Details are omitted for brevity.
(*@\textbullet{}@*) Reason step-by-step before giving your answer.Use both the text content and any available layout hints (e.g., block type, section title) to make your judgment.
(*@\textbf{\textcolor{blue}{INPUT EXAMPLE}}@*)
"The module asserts `valid` for two cycles after `ready` is high. `reset` overrides this behavior."
(*@\textbf{\textcolor{blue}{EXPECTED OUTPUT}}@*)
Modality: TEXT  
Semantic category: Timing Behavior
\end{lstlisting}

The output of the preprocessing stage is a unified intermediate representation that merges the structural information with LLM-predicted semantic annotations. In the intermediate representation, each block is encapsulated in a structured JSON-like object containing the following fields:

\begin{itemize}
    \item \texttt{content}: the raw content (text, table, or formula) extracted from the PDF block
    \item \texttt{modality}: one type of \{\texttt{TEXT}, \texttt{TABLE}, \texttt{FORMULA}, \texttt{DIAGRAM}\}, as inferred by the LLM
    \item \texttt{semantic\_category}: one of the predefined functional categories (e.g., \texttt{Timing Behavior}, \texttt{Control Logic})
    \item \texttt{page\_number}: page index of the original block
    \item \texttt{bounding\_box}: physical location of the block on the page
    \item \texttt{layout\_hints}: information from the parser (e.g., font size, block type, section ID)
\end{itemize}

This format serves as a standardized, cross-modal input format for information extraction. It ensures that semantic content and structural context are jointly preserved, enabling precise mapping of specification intent to formal representations in subsequent stages.
\subsection{Multimodal Information Extraction}
\vspace{-0.4em}
To support assertion synthesis from heterogeneous specification, AssertCoder employs modality-sensitive analyzers specialized for the modality and semantic category labels from preprocessing. Each analyzer employs a tailored prompting strategy to extract semantically relevant content. The results are then transformed into structured representations that support SVAs generation across module, signal, and temporal contexts.
\subsubsection{Text Analyzer}
Textual content blocks labeled as Interface Declaration, Control Logic, or Configuration Info contain essential semantic information about modules, ports, and their behavioral descriptions. The text analyzer processes these blocks using a structured prompt template that guides LLM to extract relevant fields, such as signal names, I/O directions, widths, and associated functionality. A representative prompt is shown in Listing~\ref{lst:text-analyzer-prompt}.
\vspace{-0.3em}
\begin{lstlisting}[caption={Prompt Template for Text Analyzer},label={lst:text-analyzer-prompt}]
(*@\texttt{\textbf{\textcolor{blue}{SYSTEM ROLE:}}}@*)
You are an expert digital hardware design analyst.
(*@\texttt{\textbf{\textcolor{blue}{TASK DESCRIPTION:}}}@*)
You will be given a paragraph from hardware specifications.  
Your job is to extract structured design information using the following YAML schema:
---
name: string
description: string
ports:
  - name: string   - I/O directions: input | output | inout
  - width: int     - description: string
implementation:
  -reset: string   - accumulation: string
  -output_behavior: string
  -counter_operation: string
example_usage: string
notes:
  - string
  - ...
module_interface: string
---
(*@\texttt{\textbf{\textcolor{blue}{INSTRUCTIONS:}}}@*)
- Use only the information present in the paragraph.
- Do not invent missing fields; leave them blank or as empty lists.
- Maintain accurate terminology and formatting.
(*@\texttt{\textbf{\textcolor{blue}{INPUT:}}}@*)
[Insert any paragraph describing a hardware module, interface, or subcomponent]
(*@\texttt{\textbf{\textcolor{blue}{OUTPUT (example):}}}@*)
name: [module_name]
description: [...]
ports:
  - name: [...]    - direction: [...]
  - width: [...]   - description: [...]
implementation:
  reset: [...]
  ...
\end{lstlisting}
\subsubsection{Diagram Analyzer}
Another modality supported by AssertCoder is graphical content, which is handled by a unified diagram analyzer. This component interprets both structural diagrams (FSMs, control/data flow) and timing diagrams by first converting them into labeled diagram representations using OCR techniques\cite{memon2020handwritten}. The semantic category identified during preprocessing determines which prompt-driven reasoning module is used to process each diagram accordingly.

Structural diagrams are mapped to control logic using pseudocode-style interpretation, while timing diagrams are converted into natural language descriptions and formal temporal constraints using operators (\texttt{F}, \texttt{G}, and \texttt{X}). This unified framework ensures consistent logic abstraction across visual modalities and supports subsequent assertion synthesis. A representative prompt used by this analyzer is shown in Listing~\ref{lst:graph-analyzer-prompt}.

\begin{lstlisting}[caption={Prompt Template for Diagram Analyzer},label={lst:graph-analyzer-prompt}]
(*@\texttt{\textbf{\textcolor{blue}{SYSTEM ROLE:}}}@*)
You are a digital hardware logic interpreter.
(*@\texttt{\textbf{\textcolor{blue}{TASK:}}}@*)
Given a diagram parsed into a graph (with nodes and edges), analyze its type and produce structured logic output.
1. If the diagram is a structural (FSM or control flow) diagram:
    - For each edge (source --(condition)--> destination), output:
        if (state == source && condition) then next_state = destination;
2. If the diagram is a waveform-based timing diagram:
    - Identify all signal transitions and timing annotations
    - Describe behavior in both:
        - Natural language
        - Temporal logic using bounded operators (F, G, X)
(*@\texttt{\textbf{\textcolor{blue}{EXAMPLE INPUT:}}}@*)
Type: Timing Diagram  
Signal Transitions:
  - ien rise at cycle 44
  - irq_flag rise at cycle 44, descend at cycle 51
(*@\texttt{\textbf{\textcolor{blue}{EXPECTED OUTPUT:}}}@*)
Natural language:
  After ien is asserted, irq_flag becomes high within 1 cycle and remains high for 6 cycles.
Temporal logic:
  ien => F[0:1] irq_flag && G[0:5] irq_flag
---
Type: FSM Diagram  
States: IDLE, REQ, GRANT  
Transitions: 
  - IDLE --(start)--> REQ  
  - REQ --(ack)--> GRANT
Pseudocode:
  if (state == IDLE && start) then next_state = REQ;
  if (state == REQ && ack) then next_state = GRANT;
\end{lstlisting}
\subsubsection{Formula and Table Analyzers}
AssertCoder also includes prompt-driven analyzers for handling formulas and tables that appear in hardware specifications. The prompt template is omitted here for brevity.

The \textit{formula analyzer} targets symbolic expressions and timing conditions presented in either natural language or mathematical notations. The extracted semantics are expressed in equations that preserve both syntactic and logical structure.

The \textit{table analyzer} focuses on structured tabular data used to specify interfaces, configuration registers, or operational modes. With layout-informed segmentation and structured prompt template, each table is parsed into a structured JSON object. These representations are serialized into a JSON array format, which integrates seamlessly with other modality outputs and facilitates unified processing in assertion generation.

\subsubsection{Merging into a unified Specification}

To support downstream assertion generation, AssertCoder integrates the outputs from all four modality-sensitive analyzers into a unified representation. For every signal or register, its relevant content is grouped together by aggregating information from multiple modalities, including structural descriptions from text and tables, behavioral logic from diagrams, and timing constraints from formulas. The following example shows the merged specification of signal \texttt{ack\_out} after multimodal analysis.

\begin{lstlisting}[language={},caption={Merged Semantic Specification for \textnormal{\texttt{ack\_out}}},label={lst:ackout_spec}]
Signal: ack_out
(*@\texttt{\textbf{\textcolor{blue}{ATTRIBUTES:}}}@*)
- Name: ack_out      - Width: 1
- Direction: output  - Default Value: 0
- Category: control signal
(*@\texttt{\textbf{\textcolor{blue}{BEHAVIORAL SEMANTICS:}}}@*)
- Control Logic: state == READ && data_valid
- FSM Transition: READ --(data_valid)--> SEND_ACK
- Timing Constraint: ##[1:2] ack_out[*2]
- Temporal Logic: F[1:2] ack_out && G[0:1] ack_out
- Natural Language: 
  "If data_valid is asserted during READ, 
   then ack_out must be asserted within 2 cycles 
   and held for at least 1 cycle."
(*@\texttt{\textbf{\textcolor{blue}{TEMPORAL ROLES:}}}@*)
- responder   - bounded-delay   - stabilizer
(*@\texttt{\textbf{\textcolor{blue}{INTENT CANDIDATE TRIPLET:}}}@*)
- Precondition: state == READ && data_valid
- Consequence: ack_out == 1
- Timing: ##[1:2] ack_out[*2]
(*@\texttt{\textbf{\textcolor{blue}{SOURCE TRACEABILITY:}}}@*)
- Text Segment: Section 2.1 
- FSM Diagram: Figure 3
- Timing Waveform: Figure 5
- Formal Formula: Equation (2)
- Interface Table: Table 1
\end{lstlisting}

This integrated format provides a complete and structured view of each design, making it available for the subsequent assertion generation phase.

\subsection{SVA Generation}
Given the structured specification produced in the previous stage, AssertCoder generates SVAs through a multi-step CoT prompting strategy, which systematically decomposes the generation task into interpretable sub-steps. For signals of each module, the LLM-guided process performs the following four steps: (1) \textbf{Unified Semantic Decomposition}, (2) \textbf{Pattern Selection}, (3) \textbf{Temporal Binding}, and (4) \textbf{SVA Syntax Synthesis}. The details can be found in Listing\ref{lst:sva-generation-prompt}.

To perform the four-step method, AssertCoder employs a prompt template that guides each reasoning step, enabling interpretable generation and closer alignment with formal semantics. Domain knowledge is enhanced via a  RAG tool\cite{guo2025lightragsimplefastretrievalaugmented}, which injects relevant passages from a curated corpus of SVA references and hardware design literature.

The generation interface also supports feedback from the evaluation stage through undetected mutation points. While the initial synthesis is performed without feedback, subsequent iterations incorporate mutation-specific cues (mutation points) using the same prompting structure to improve mutation detection. 

\begin{lstlisting}[caption={Prompt Template for SVA Generator},label={lst:sva-generation-prompt}]
(*@\textbf{\textcolor{blue}{SYSTEM ROLE:}}@*)
You are a hardware verification expert specialized in SVA generation.
(*@\textbf{\textcolor{blue}{TASK DESCRIPTION:}}@*)
Given a structured specification segment for a signal (including precondition, consequence, and timing window), generate a valid SVA in four steps:
[Optional]
If mutation detection feedback is available, consider the following uncovered mutation conditions to guide assertion refinement:
- mutation_points: ["read_valid never asserted", "read_req never followed by output", ...]
(*@\textbf{\textcolor{blue}{Step 1 — Semantic Decomposition:}}@*)
Extract the intent into three parts:
- Precondition: When should this assertion be triggered?
- Consequence: What signal behavior must happen?
- Timing: Within what time window must it occur?
(*@\textbf{\textcolor{blue}{Step 2 — Pattern Selection:}}@*)
Choose the best matching assertion pattern:
- Implication (if A then B within N cycles)
- Stability (signal must remain constant)
- Invariant (property must always hold)
(*@\textbf{\textcolor{blue}{Step 3 — Temporal Binding:}}@*)
Translate timing into SystemVerilog syntax, e.g., ##[2:4], [*2], etc.
(*@\textbf{\textcolor{blue}{Step 4 — SVA Syntax Synthesis:}}@*)
Assemble the final assertion using SystemVerilog syntax. Include clock edge and proper operators.
(*@\textbf{\textcolor{blue}{INPUT EXAMPLE:}}@*)
{
  "precondition": "read_req is high && arbiter_grant is low",
  "consequence": "read_valid must be high",
  "timing": "between 2 and 4 cycles, and hold for >=2 cycles"
  "mutation_points": []  // leave empty during initial generation
}
(*@\textbf{\textcolor{blue}{EXPECTED OUTPUT:}}@*)
Step 1:
If (read_req && !arbiter_grant), then read_valid must become high within 2-4 cycles and stay high for >=2 cycles.
Step 2:Pattern: Implication
Step 3:Temporal: ##[2:4] read_valid[*2]
Step 4:assert property (@(posedge clk)(read_req && !arbiter_grant) |-> ##[2:4]read_valid[*2]);
\end{lstlisting}
\subsection{Evaluation and Iteration}
After the initial round of assertion generation, AssertCoder evaluates the effectiveness of the SVA set through mutation injection and model checking. Let $A = \{a_1, a_2, \dots, a_n\}$ be the set of generated assertions and $M = \{m_1, m_2, \dots, m_k\}$ the set of injected mutations. For each assertion-mutant pair $(a_i, m_j)$, model checking is performed, and the outcome is logged as:
\begin{equation}
\text{detect}(a_i, m_j) =
\begin{cases}
1 & \text{if $a_i$ fails on $m_j$} \\
0 & \text{otherwise}
\end{cases}
\end{equation}

We define the \textbf{mutation Score} for each assertion $a_i$ as:
\begin{equation}
\text{Score}(a_i) = \sum_{j=1}^{k} \text{detect}(a_i, m_j)
\end{equation}

This score represents the number of unique mutations that $a_i$ is able to detect. Based on this metric, to further assess the overall effectiveness of the assertion set, we compute the \emph{Average Mutation Score} as:
\begin{equation}
\text{Avg. Mutation Score} = \frac{1}{n} \sum_{i=1}^{n} \text{Score}(a_i)
\end{equation}

This metric reflects the average number of detected mutations per assertion, capturing the aggregate detection strength of the generated set. Assertions with $\text{Score}(a_i) = 0$ are deemed low-quality and automatically removed from the final set. To evaluate overall detection, we define the \textbf{Mutation detection Rate (MDR)}:
\begin{equation}
\text{MDR} = \frac{\left| \left\{ m_j \,\middle|\, \exists a_i : \text{detect}(a_i, m_j) = 1 \right\} \right|}{k}
\end{equation}
\begin{table*}[t]
\scriptsize
\centering
\caption{ASSERTION GENERATION QUALITY COMPARISONS ACROSS THREE DESIGNS}
\label{tab:assertion_comparison}
\begin{tabular}{lccccccc}
\toprule
\textbf{Design} & \textbf{Method} & \textbf{\#SVAs Gen.} & \textbf{Syntax Correctness (\%)} & \textbf{Functional Correctness (\%)} & \textbf{Avg. Mutation Score} & \textbf{MDR (\%)} & \textbf{FPR (\%)} \\
\midrule
\multirow{3}{*}{I2C}
& AssertCoder & \textbf{92} & \textbf{90 (97.8\%)} & \textbf{81 (88.0\%)} & \textbf{0.63} & \textbf{84.3} & \textbf{3.2} \\
& AssertLLM & 89 & 86 (96.6\%) & 73 (82.0\%) & 0.59 & 79.8 & 5.6 \\
& GPT-4o & 112 & 68 (60.7\%) & 22 (19.6\%) & 0.17 & 41.0 & 19.3 \\
\midrule
\multirow{3}{*}{AES}
& AssertCoder & \textbf{134} & \textbf{132 (98.5\%)} & \textbf{118 (88.1\%)} & \textbf{0.61} & \textbf{87.2} & \textbf{2.8} \\
& AssertLLM & 130 & 128 (98.5\%) & 102 (78.5\%) & 0.57 & 81.5 & 4.9 \\
& GPT-4o & 141 & 74 (52.5\%) & 28 (19.9\%) & 0.16 & 38.5 & 21.1 \\
\midrule
\multirow{3}{*}{Openmsp430}
& AssertCoder & \textbf{215} & \textbf{209 (97.2\%)} & \textbf{187 (86.9\%)} & \textbf{0.59} & \textbf{85.4} & \textbf{3.5} \\
& AssertLLM & 210 & 205 (97.6\%) & 165 (78.6\%) & 0.54 & 78.1 & 6.1 \\
& GPT-4o & 236 & 101 (42.8\%) & 34 (14.4\%) & 0.13 & 34.7 & 24.8 \\
\midrule
\textbf{Avg.}
& AssertCoder & \textbf{147.0} & \textbf{143.7 (97.8\%)} & \textbf{128.7 (87.6\%)} & \textbf{0.61} & \textbf{85.6} & \textbf{3.2} \\
& AssertLLM & 143.0 & 139.7 (97.7\%) & 113.3 (79.2\%) & 0.57 & 79.8 & 5.5 \\
& GPT-4o & 163.0 & 81.0 (49.7\%) & 28.0 (17.2\%) & 0.15 & 38.1 & 21.7 \\
\bottomrule
\end{tabular}
\vspace{-20pt}
\end{table*}
If $\text{MDR} \textless 1$, it implies that certain mutations remain undetected by all generated assertions. In such cases, the mutation traces and logs serve as feedback to regeneration SVA. The failed mutations’ context (e.g., affected signal, location, or operation) is used to promote the LLM to produce more relevant assertions.

By employing model checking, each assertion is re-evaluated against the unmutated DUT to identify potential false positives. Assertions that are syntactically correct but semantically misaligned with the specification, producing spurious counterexamples on the original DUT, are labeled as false positives. Those that fail due to detecting actual DUT flaws are not considered false positives and may even motivate specification updates. We define the \textit{False Positive Rate (FPR)} as:
\begin{equation}
\scriptsize
\text{FPR} = \frac{|\{a_i \in A_{\text{final}} \mid \texttt{fail}(a_i, \text{DUT}) \wedge \neg \texttt{semantically\_corr}(a_i)\}|}{|A_{\text{final}}|}
\end{equation}
where:
\begin{itemize}[topsep=0pt, partopsep=0pt, parsep=0pt, itemsep=0pt]
    \item $A_{\text{final}}$ denotes the final set of assertions after filtering and regeneration;
    \item $\texttt{fail}(a_i, \text{DUT})$ indicates assertion $a_i$ fails in model checking on the DUT;
    \item $\texttt{semantically\_corr}(a_i)$ is true if $a_i$ aligns with specification intent.
\end{itemize}

Since assertion failures may indicate either flawed assertions or actual DUT bugs, we perform a post-validation step that manually checks failed assertions. 
\vspace{-2pt}
\section{Experimental Evaluation}
\vspace{-0.5em}
\subsection{Experimental Setup}
\vspace{-0.5em}
We evaluate \textbf{AssertCoder} on a suite of open-source hardware designs of varying complexity, including an \textit{I2C master}\cite{herveille20032} (small-scale), an \textit{AES core}\cite{hsing2013aes} (medium-scale), and  \textit{Openmsp430}\cite{girard2009openmsp430} (large-scale). Each design is accompanied by its multimodal specification document, containing natural language descriptions, tables, and timing diagrams.

The proposed method is compared against two approaches: (1) \textbf{GPT-4o}\cite{openai2024gpt4}, used in a direct zero-shot prompting manner, and (2) \textbf{AssertLLM}, a recent multi-agent prompting framework for assertion synthesis. We employ AssertLLM as the baseline method. All methods operate on identical specification inputs. Our framework uses the GPT-4o API for generation, SymbiYosys\cite{wolf2016yosys} for model checking, and Mantra\cite{wu2023mantra} for injecting mutations based on bug-derived mutation operators. Approximately 100--300 mutants are generated per DUT.
\vspace{-4pt}
\subsection{Evaluation of Assertion Quality}
\vspace{-0.5em}
To assess assertions quality, we analyze the outputs of \textbf{AssertCoder}, \textbf{AssertLLM}, and \textbf{GPT-4o} on three DUTs using a set of standard metrics. Table~\ref{tab:assertion_comparison} summarizes the results.

Across all designs, AssertCoder achieves the highest assertion quality. It maintains a very high accuracy rate in both syntax correctness and functional correctness, and generates more effective assertions. While AssertLLM demonstrates strong effectiveness, it occasionally produces semantically weak assertions due to insufficient temporal grounding. In contrast, GPT-4o struggles to produce useful SVAs in the absence of structural guidance, yielding a low mutation detection rate and high false positive rate.

The advantage of AssertCoder lies in its unified semantic decomposition and CoT reasoning pipeline, which enables precise identification of assertion intent and timing constraints. Importantly, the low FPR indicates that the SVA generated by the AssertCoder framework has high quality. These results demonstrate the importance of structured prompting and multimodal signal-level abstraction in SVA generation.
\vspace{-0.5em}
\subsection{Ablation Study}
\vspace{-4pt}
To understand the contribution of each core component in the AssertCoder pipeline, we conduct an ablation study by incrementally disabling three major modules: (1) \textit{RAG}, (2) \textit{CoT prompting} , and (3) \textit{mutation-guided iterative refinement}. We measure the impact on both the number of functional correctness assertions and the MDR across all three DUTs.

\begin{figure}[t]
    \centering
    \includegraphics[width=0.85\linewidth]{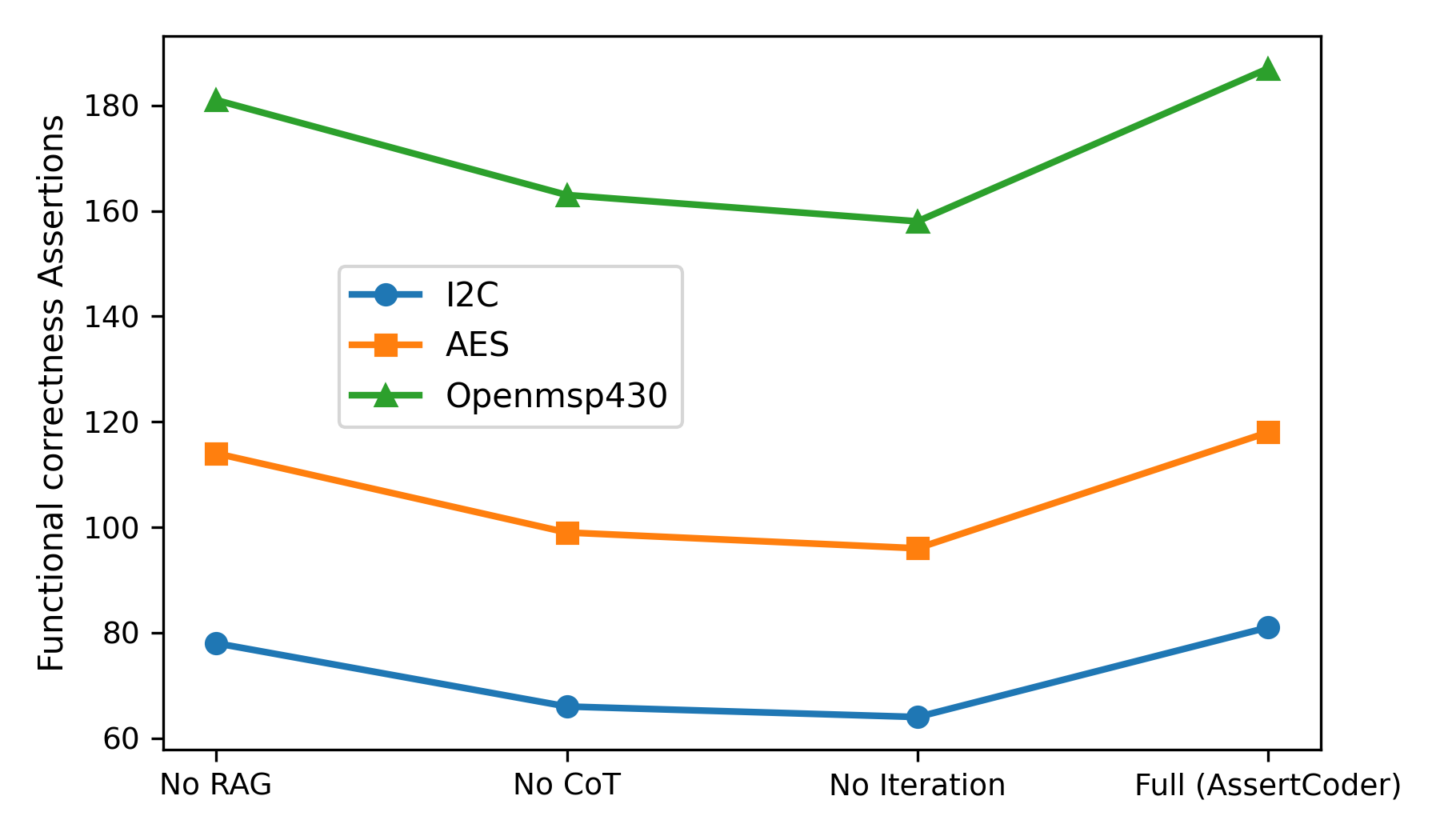}
    \caption{Effect of component ablations on functional correctness assertions.}
    \label{fig:ablation_valid}
\end{figure}

\begin{figure}[t]
    \centering
    \includegraphics[width=0.85\linewidth]{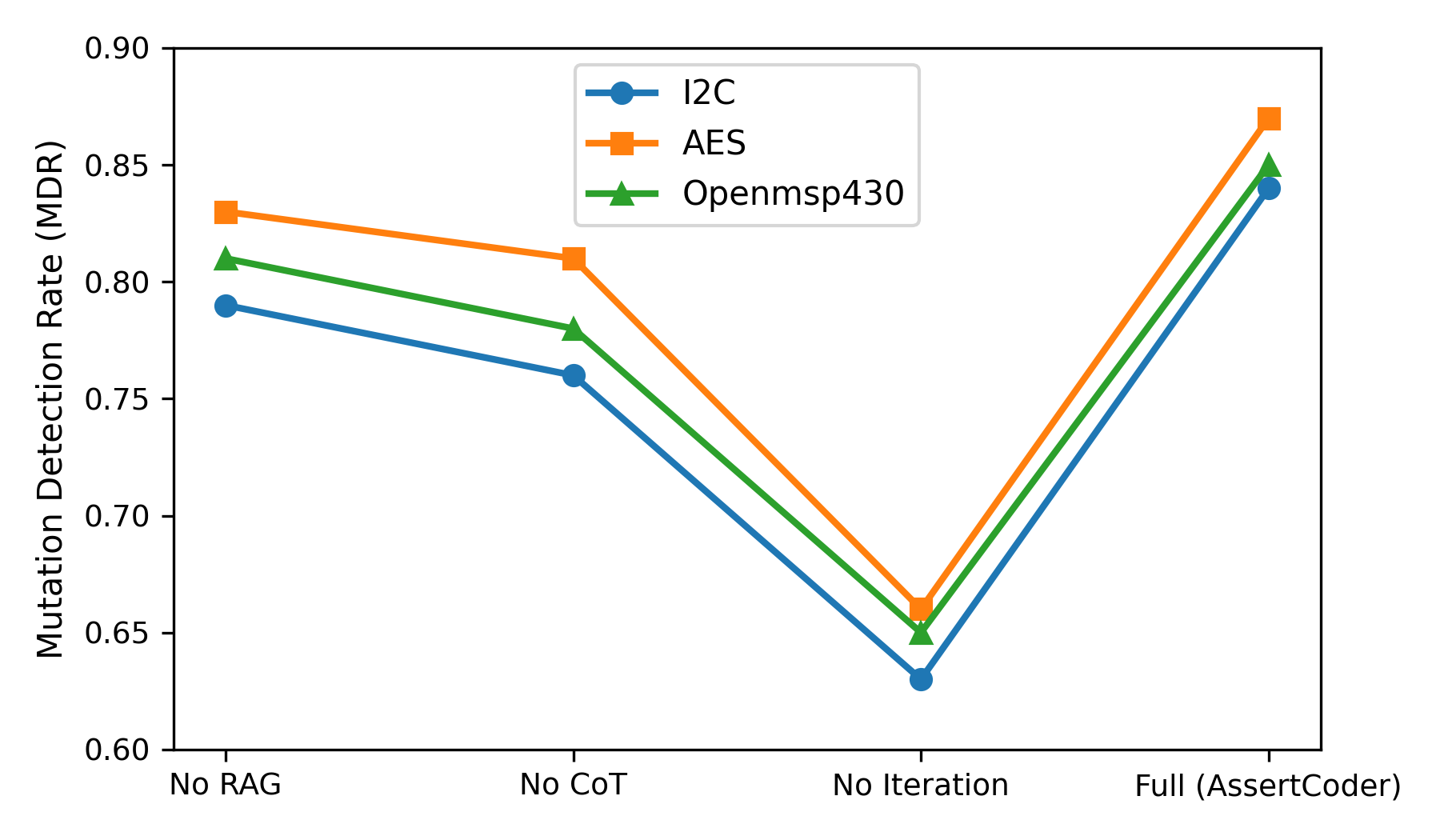}
    \caption{Impact on mutation detection rate (MDR) across designs.}
    \label{fig:ablation_mdr}
\end{figure}

As shown in Figure~\ref{fig:ablation_valid}, removing CoT prompting leads to the largest decrease in correctness SVA count, highlighting its importance in guiding multi-step reasoning and ensuring signal-condition alignment. Removing RAG has comparatively minimal impact, indicating that most assertion-relevant semantics are already encoded within the specification. However, without iterative refinement, the system produces fewer high-impact assertions and achieves a lower MDR.

Figure~\ref{fig:ablation_mdr} further confirms that iterative assertion synthesis is critical for maximizing mutation detection, as its removal leads to substantial detection drops across all DUTs. These results confirm that all three components contribute to AssertCoder's effectiveness, with CoT enabling correctness and iteration ensuring completeness.
\vspace{-0.5em}
\section{Conclusion}
In this work, we introduced AssertCoder, a unified framework for automatic SVA generation from multimodal hardware specifications. By integrating modality-sensitive semantic extraction, a structured multi-step Chain-of-Thought prompting strategy, and mutation-guided iterative refinement, AssertCoder significantly improves the quality of assertion-based verification. Experimental evaluations demonstrated that AssertCoder outperforms existing methods, achieving higher functional correctness and mutation detection.
\bibliographystyle{IEEEtran}  
\bibliography{main}           
\end{document}